\documentclass[reprint,apl,superscriptaddress]{revtex4-1}

\usepackage{hyperref,graphicx,textcomp}
\usepackage{amsmath}

\makeatletter
\renewcommand\@biblabel[1]{#1.}
\makeatother

\begin{abstract}
  Solution-processed amorphous silicon is a promising material for
  semiconductor devices.  Unfortunately, its manufacturing leaves a high
  density of defects in the layer, which can be reduced by a treatment with
  hydrogen radicals.  Here, we present an optimized hydrogen treatment, which
  is used for best performing solar cells made of solution-processed amorphous
  silicon.  We examine the amount and the nature of hydrogen incorporation
  using infrared absorption and hydrogen effusion.  The hydrogen treatment not
  only increases hydrogen content significantly, it also enlarges the fraction
  of hydrogen in a bonding configuration which is known to be advantageous for
  electronic properties, albeit only close to the surface.  Using electron spin
  resonance and and photothermal deflection spectroscopy spectra, we confirm a
  reduction of defect density.  Regarding the electrical properties, the ratio
  of photo and dark conductivity is increased by almost two decades.  This
  leads to a greatly enhanced performance of solar cell devices which use the
  material as the absorber layer.  In particular, the efficiency jumps by a
  factor of three.
\end{abstract}

\begin{document}

\title{Influence of hydrogen radicals treatment on layers and solar cells made
  of solution-processed amorphous silicon}

\author{Torsten Bronger}
\affiliation{Forschungszentrum J\"ulich GmbH, Institut f\"ur Energie und
  Klimaforschung (IEK-5), J\"ulich, 52425 (Germany)}
\author{Jan W\"ordenweber}
\affiliation{Forschungszentrum J\"ulich GmbH, Institut f\"ur Energie und
  Klimaforschung (IEK-5), J\"ulich, 52425 (Germany)}
\author{Paul W\"obkenberg}
\affiliation{Evonik Industries AG Paul-Baumann Str.~1, Marl, 45772 (Germany)}
\author{Stefan Muthmann}
\affiliation{Forschungszentrum J\"ulich GmbH, Institut f\"ur Energie und
  Klimaforschung (IEK-5), J\"ulich, 52425 (Germany)}
\author{Odo Wunnicke}
\affiliation{Evonik Industries AG Paul-Baumann Str.~1, Marl, 45772 (Germany)}
\author{Reinhard Carius}
\affiliation{Forschungszentrum J\"ulich GmbH, Institut f\"ur Energie und
  Klimaforschung (IEK-5), J\"ulich, 52425 (Germany)}


\maketitle


Since the work of Shimoda\cite{shimoda2006solution}, solution-processed
amorphous silicon made of cyclopentasilane (CPS) and neopentasilane (NPS)
precursors has gained interest for the application in semiconductor devices, in
particular solar cells.  Recently, solar cells based on this material reached
an efficiency of 3.5\,\%~\cite{bronger2014solution}.  However, immediately
after the conversion of the polymer into a silicon layer, layer quality is
poor.\cite{masuda2012fabrication} Thus, a treatment with hydrogen radicals is
applied to bring hydrogen back into the
layer.\cite{masuda2012fabrication,bronger2014solution,sontheimer2014solution}
Since this is a crucial step in the processing, it is worth having a close look
at it.



For details of layer and stack preparation, including the doping of the
material, see the experimental section in
Ref.~\onlinecite{bronger2014solution}.


For the hydrogen radicals treatment, the sample is mounted on a holder in a
vacuum chamber (approx.~$4\cdot10^{-7}$\,mbar).  It is tempered for one hour at
370\,\textcelsius, so that the heat is homogeneously distributed and the
surface water film evaporated.  Then, 30\,sccm of hydrogen gas (chamber
pressure is then at 0.1\,mbar) pass the layer.  At the same time, a tantalum
filament at 1350\,\textcelsius\ with a distance of 6.8\,mm to the sample
decomposes the hydrogen.  The sample is heated with the same power as in the
tempering step, however the filament would have an additional thermal impact,
which we have not measured.  This passivation step takes 2~hours, after which
the sample is thermalized to room temperature and transferred out of the
chamber into ambient atmosphere.

A dip of the sample in hydroflouric acid (10\,\%) ca.\ 3~minutes before the
transfer into vacuum does not improve layer or cell performance.  However, it
is very important to check the chamber wall for deposits.  We observed black
discoloration on the wall after a couple of hydrogen treatment runs, which
negatively affected passivation effectiveness.


\begin{figure}
  \includegraphics[width=3in]{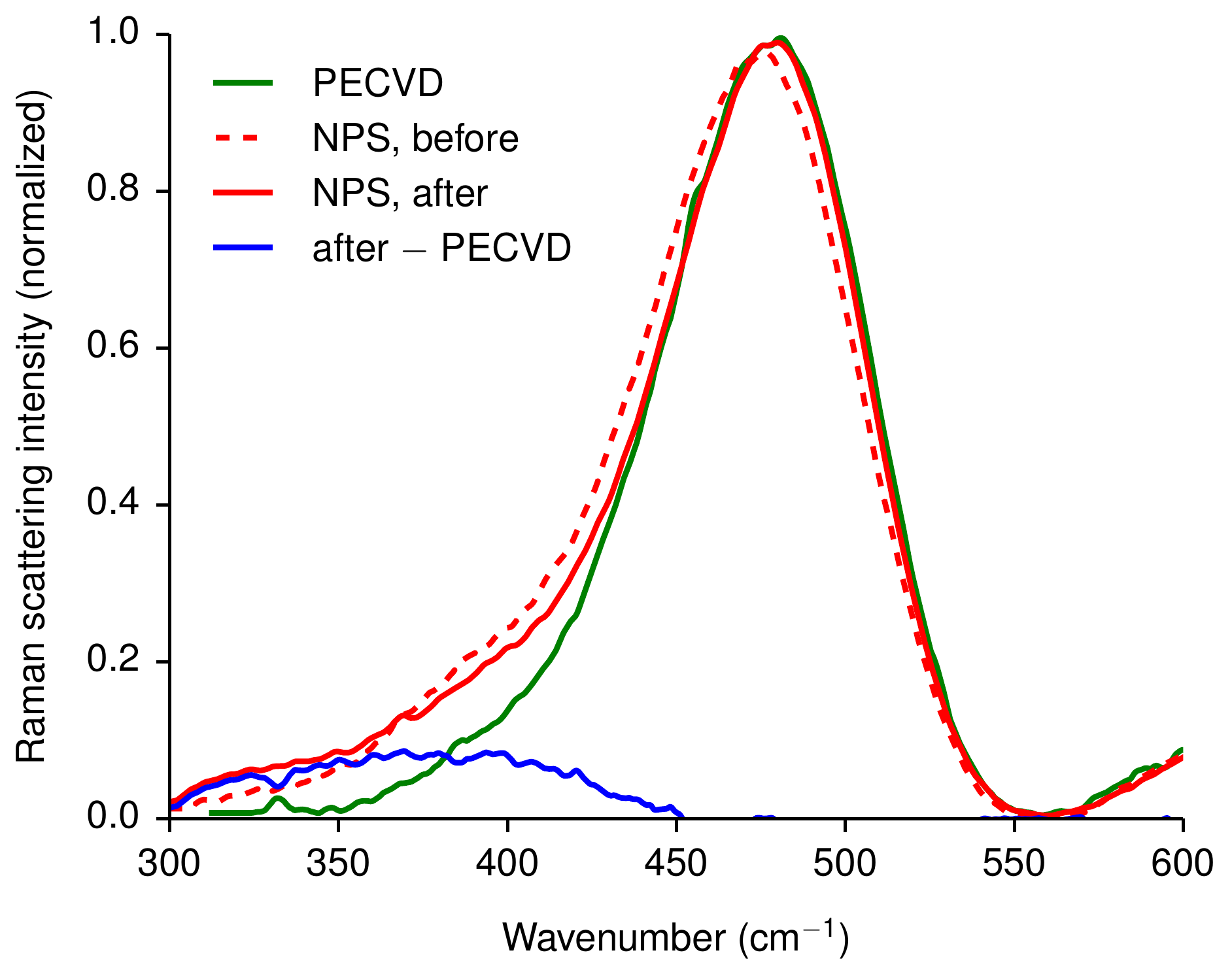}
  \caption{Spectrum of Raman scattering of intrinsic NPS and
    state-of-the-art PECVD material.
    The amorphous silicon peaks are normalized to~1.}
  \label{fig:raman}
\end{figure}

\begin{table*}
  \newlength{\zellenbreite}
  \settowidth{\zellenbreite}{$-9.9$}
  \caption{Electrical and optical properties of NPS material at room
    temperature.  The optical gap is the energy at which the absorption
    coefficient equals $10^{-4}\,\mathrm{cm}^{-1}$, as determined by the PDS
    technique.  Typical values for state-of-the-art PECVD material are shown
    for comparison.}
  \label{tab:layer-properties}
  \begin{ruledtabular}
  \begin{tabular}{@{}lccc@{}}
    &before & after & PECVD \\
    &\multicolumn{2}{c}{hydrogen treatment} & reference \\
    Raman peak shift (cm$^{-1}$) & $-6.8$ & $-2.7$ & \phantom00\phantom{.0} \\
    Raman peak width (cm$^{-1}$) & \hbox to \zellenbreite{\hfill$75.3$} & \hbox to \zellenbreite{\hfill$72.2$} & $70.8$ \\
    \hline
    optical gap (eV) & 1.95 & 1.99 & 1.98 \\
    optical band tail width (meV) & 86 & 78 & 61 \\
    absorption at 1.2\,eV (cm$^{-1}$) & 24 & 7.0 & 4 \\
    \hline
    hydrogen content (\%) & 6.9 & 8.4 & 17\\
    micro structure factor (\%) & 64 & 52 & 22\\
    \hline
    dark conductivity (S/cm) & $2.2\cdot10^{-10}$ & $3.3\cdot10^{-11}$ & $1.4\cdot10^{-10}$ \\
    photo conductivity (S/cm) & $2.2\cdot10^{-7\phantom0}$ & $2.5\cdot10^{-6\phantom0}$ & $3.4\cdot10^{-5\phantom0}$ \\
    photo/dark conductivity ratio & $1.0\cdot10^{3}$ & $7.7\cdot10^{4}$ & $2.4\cdot10^{5}$ \\
  \end{tabular}
  \end{ruledtabular}
\end{table*}

The Raman specra are measured with an excitation wavelength of 488\,nm.  Our
in-house design is capable of a resolution of $0.5\,\mathrm{cm}^{-1}$.
Moreover, we used a widened laser spot of $50\times1\,\text{\textmu m}$ in size
in order to be able to average over layer inhomogeneities.

Fig.~\ref{fig:raman} shows the Raman scattering spectrum of intrinsic layers
fabricated of NPS\@.  The dominating peak is that of amorphous silicon with no
measurable crystalline volume fraction.  In contrast to PECVD material, the
peak is slightly shifted towards smaller wavenumbers, and broadened.
Tab.~\ref{tab:layer-properties} quantifies these differences.  These two
deviations from the PECVD reference spectrum are largely compensated by the
treatment with hydrogen radicals, as visualized in Fig.~\ref{fig:raman} and
shown in Tab.~\ref{tab:layer-properties}\@.  The residual difference is plotted
in blue in Fig.~\ref{fig:raman}.

One may deduce from these results that the hydrogen treatment reduces the
stress.  This is backed by the peeling-off of very thick layers.  Thus, we know
that it suffers from tensile stress.  However, this explanation is put in
question by crack formation and propagation being independent of the hydrogen
treatment.  Moreover, the mechanism of stress reduction is unclear.  An
alternative explanation for the shift would be a peak at 460\,cm$^{-1}$ which
vanishes or is strongly suppressed after the treatment, but also here, the
mechanism would be unclear.  Future substrate curvature measurements as an
estimate for the layer tension may provide further insight.

The residual difference between hydrogen-treated NPS material and PECVD
material (blue in Fig.~\ref{fig:raman}) can be easily explained by an
enhanced longitudinal optical (LO) phonon absorption at 380\,cm$^{-1}$ in
NPS\@.  Generally, this indicates increased disorder in the
microstructure.\cite{gerbi2003increasing}

\begin{figure}
  \includegraphics[width=3in]{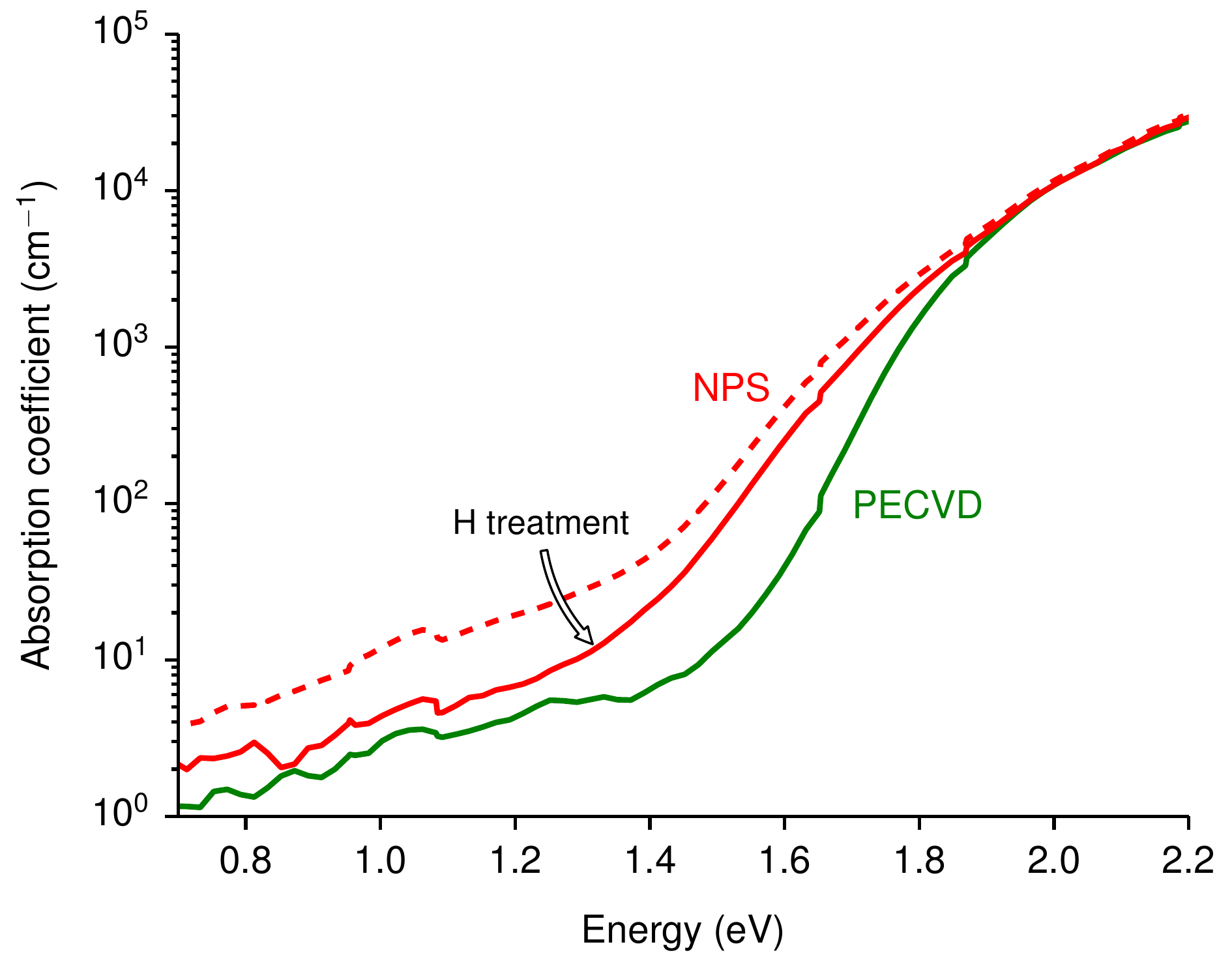}
  \caption{The absorption coefficient versus photon energy of intrinsic layers
    of NPS and PECVD material as measured with the PDS technique.  The
    thickness of the NPS sample is approximately 134 nm, while that of the
    PECVD sample is 200 nm.}
  \label{fig:pdf}
\end{figure}

For the absorption measurements, we use a photothermal deflection spectroscopy
(PDS) in-house design.  A cuvette filled with carbon tetrachloride takes the
sample.  Our monochromator after a hydrogen lamp allows an output bandwidth of
10~meV, however, the data points are spaced by 20~meV\@.  The integral
irradiation power on the sample is approx.\ 1~\textmu W\@.

Fig.~\ref{fig:pdf} depicts the absorption spectrum of an intrinsic NPS layer,
measured with the PDS technique.  NPS exhibits a strongly enhanced absorption
in the critical domain below~2\,eV\@.  The optical tail width, which we define
as the maximal slope of the curve in semi-log\-a\-rith\-mic axes, is
substantially higher than in the PECVD case.  The same is true for the
absorption at 1.2\,eV\@.  Similarly to the Raman results, the hydrogen
treatment diminishes the differences between the NPS layer and the reference
layer.  Tab.~\ref{tab:layer-properties} confirms this observation
quantitatively.  Additionally, it gives the value of the optical gap for all
cases.  Again, the hydrogen treatment brings this value closer to the PECVD
reference.

\begin{figure}
  \includegraphics[width=3in]{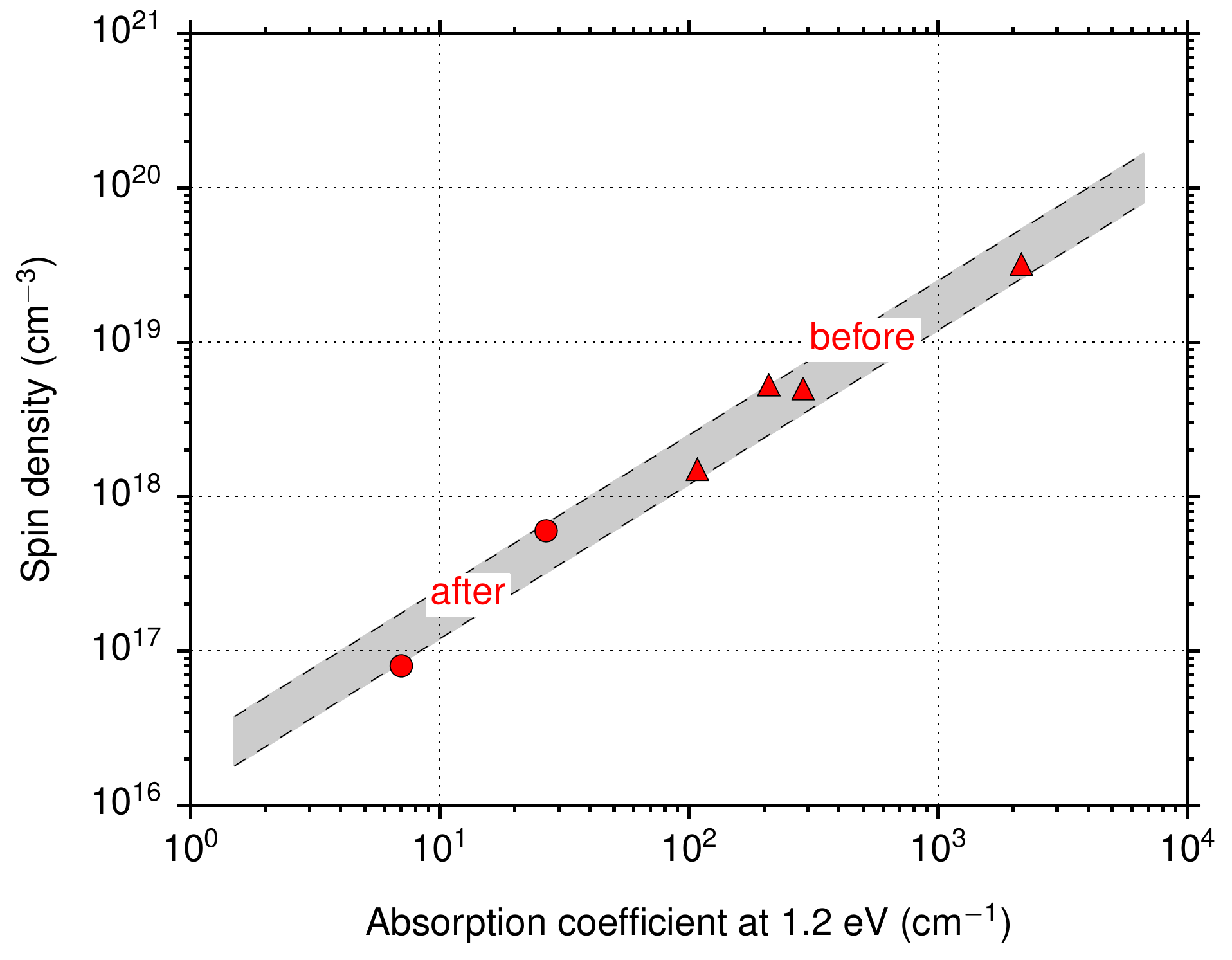}
  \caption{Dependency of spin density on optical absorption at 1.2\,eV\@.
    Samples before the hydrogen treatment (red triangles) as well as after (red
    discs) are included.  The gray corridor is taken from
    Ref.~\onlinecite{wyrsch1991reach} and bases on a comprehensive results
    compilation.}
  \label{fig:esr_pdf}
\end{figure}

The hydrogen treatment saturates the dangling bonds, thus the absoption at
1.2\,eV is reduced.  This is confirmed strongly by the very good correlation of
PDS and ESR results (see Fig.~\ref{fig:esr_pdf}).  Furthermore, the decrease in
tail width is interpreted as less disorder in microstructure, which matches
nicely the lower LO phonon absorption in Raman scattering.  This also leads to
the increased optical gap as defined in Tab.~\ref{tab:layer-properties}.

Conventional continuous wave electron spin resonance (ESR) measurements are
performed with a commercial X-band ($\nu=9.3$~GHz) Bruker Elexsys E500
spectrometer in a cylindrical mode resonator at room temperature, microwave
power of $4\cdot10^{-5}$~W, magnetic field modulation amplitude of 5~G, and
modulation frequency of 100~kHz.  A calibrated sputtered amorphous silicon
sample with a spin density of $2\cdot10^{15}\,\mathrm{cm}^{-3}$ and g-value of
2.00565 was used as a reference standard.

Fig.~\ref{fig:esr_pdf} combines the results of PDS and ESR measurements.  The
optical absorption at 1.2\,eV is proportional to the defect density as measured
with ESR,\cite{jackson1982direct,wyrsch1991reach} and indeed this dependency is
reproduced by NPS samples with high accuracy over 2.5 orders of magnitude.
Ref.~\onlinecite{wyrsch1991reach} reports a proportionality factor in the range
1.2--$2.5\cdot10^{16}\,\mathrm{cm}^{-2}$, which is shaded in gray in
Fig.~\ref{fig:esr_pdf}.  Obviously, the NPS samples both before and after the
hydrogen treatment stay in this range.

\begin{figure}
  \includegraphics[width=3in]{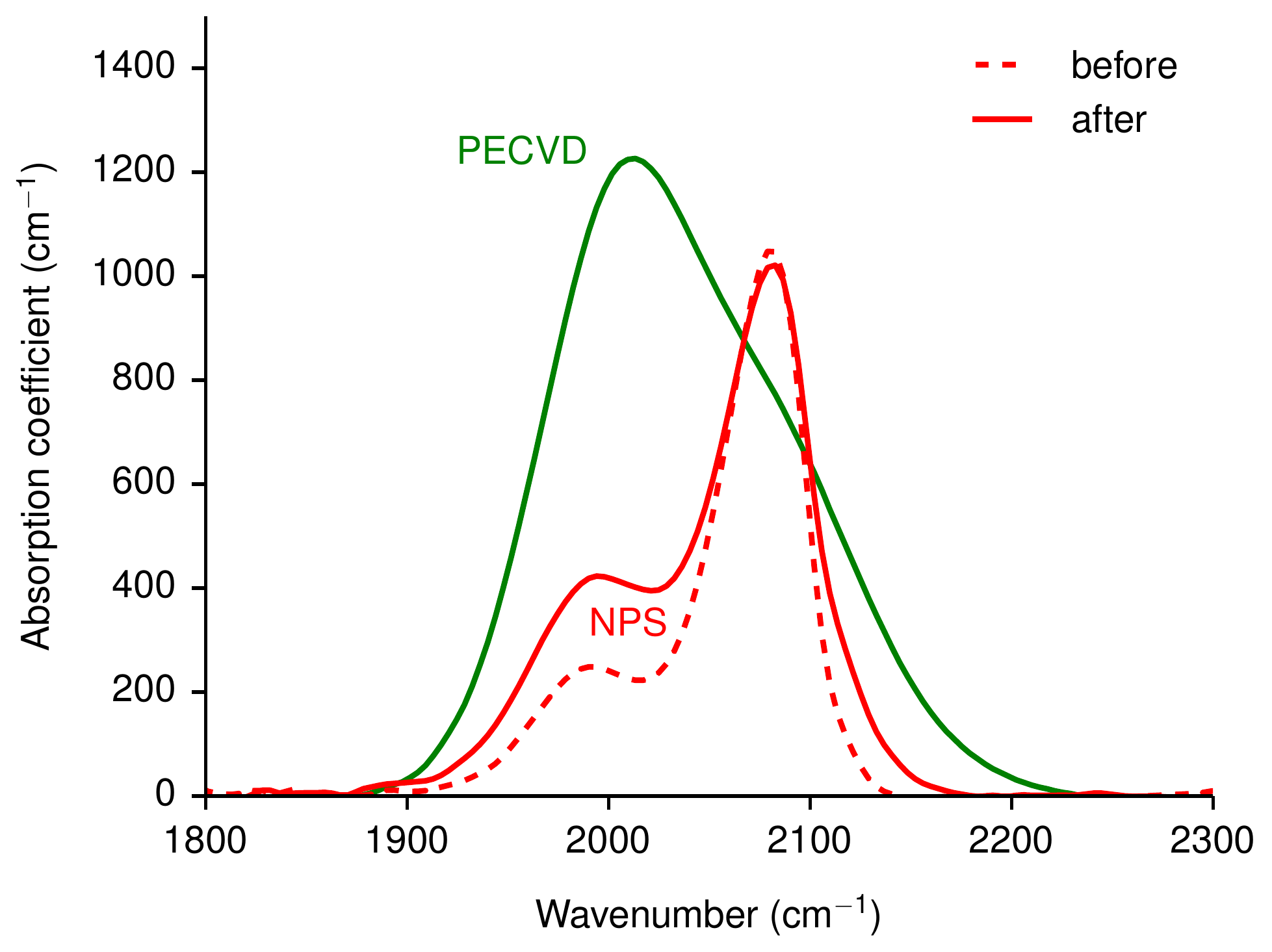}
  \caption{Infrared absorption spectrum of intrinsic NPS (230 nm) and PECVD
    (700 nm) layers, deposited on silicon wafer substrate.  The dashed line
    denotes material before the hydrogen treatment.  This plot is taken from
    the supplementary information of Ref.~\onlinecite{bronger2014solution}.}
  \label{fig:ir}
\end{figure}

For the infrared absorption measurements, we deposited the NPS layer on
double-side polished silicon wafer.  The measurement data itself is collected
by an FTIR spectrometer (Nicolet 5700) with a glow bar as light source.  We
substracted the absorption spectrum of a piece of uncovered silicon wafer, and
substracted an additional, manually tweaked spline baseline.

Fig.~\ref{fig:ir} shows infrared absorption spectrum of an NPS layer.  There
are two peaks of interest here, one at approx.\ 2000\,cm$^{-1}$ and the other
at approx.\ 2080\,cm$^{-1}$.  Both are identified with \mbox{Si--H} oscillation
modes, however of hydrogen in different bonding configurations.  In particular,
the 2080 peak is related to bonds at (inner)
surfaces,\cite{cardona1983vibrational} which is not able to actually passivate
dangling bonds in the bulk.  In contrast, the 2000 peak is considered ``good''
hydrogen in the sense of decreasing bulk defect density.  Quantitatively, this
is expressed by the microstructure factor (MSF).\cite{mahan1987evidence}
Generally speaking, a smaller MSF indicates a better material.

As one can see in Tab.~\ref{tab:layer-properties}, in NPS layers, the MSF is
significantly higher than in PECVD reference layers.  One can also see that the
hydrogen treatment ameliorates this situation to some extent.
Fig.~\ref{fig:ir} shows that this is achieved by increasing the bulk hydrogen
while leaving the hydrogen at inner surfaces constant.  Accordingly, the
integral hydrogen content raises by 20\,\%.  However, the content of
passivating hydrogen is still much lower than in PECVD material.

\begin{figure}
  \includegraphics[width=3in]{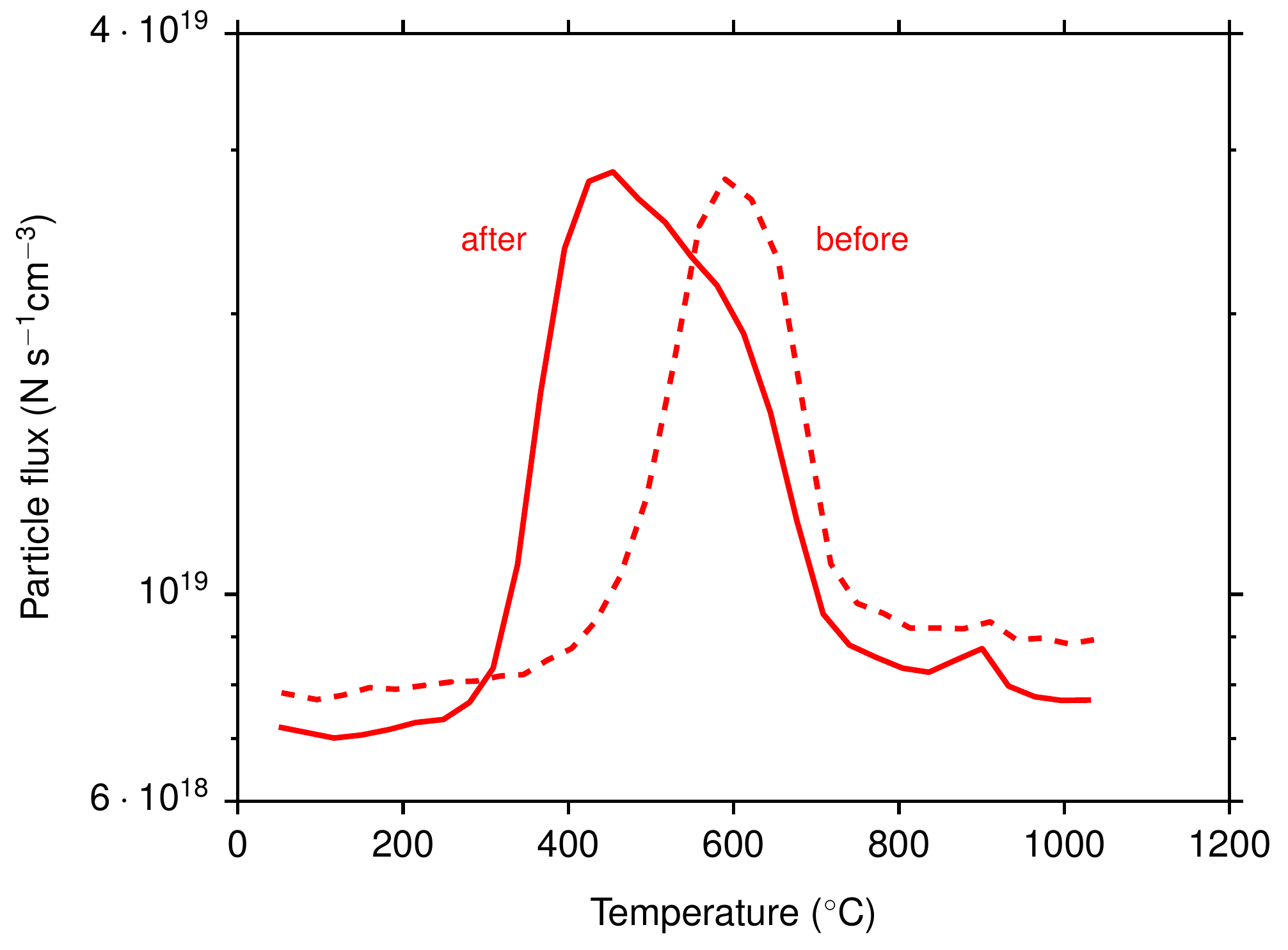}
  \caption{Hydrogen effusion diagram of an intrinsic NPS layer before and after
    the hydrogen treatment.}
  \label{fig:effusion}
\end{figure}

\begin{table*}
  \caption{Solar cell properties at room
    temperature.  Typical values for state-of-the-art PECVD material are shown
    for comparison.}
  \label{tab:cell-properties}
  \begin{ruledtabular}
  \begin{tabular}{@{}l@{}cccccc@{}}
    & effective cell & efficiency & $I_{sc}$ in & $V_{oc}$ & fill factor &
    series resistance\\
    & area in mm$^2$ & in \%      & mA\,cm$^{-2}$ & in mV    & in \% & in Ohm\\
    before H treatment & 3.2 & 0.68 & 4.4 & 497 & 31.1 & 2,400\\
    after H treatment & 3.2 & 2.0 & 7.4 & 628 & 42.7 & \phantom{0,}910 \\
    \hline
    best NPS cell\cite{bronger2014solution} & 3.2 & 3.5 & 9.0 & 730 & 53.8 &
    \phantom{0,}550\\
    PECVD\cite{bronger2014solution} & 3.2 & 5.9 & 9.8 & 892 & 67.5 &
    \phantom{0,}450 \\
  \end{tabular}
  \end{ruledtabular}
\end{table*}

These findings are accompanied by hydrogen effusion measurements that are
presented in Fig.~\ref{fig:effusion}.  Hydrogen effusion is measured in the
setup described in Ref.~\onlinecite{beyer2011hydrogen}, a so-called open system
with a turbomolecular pump.  The base pressure is approx.\ $10^{-10}$\,mbar and
the heating rate 20\,\textcelsius/min.  NPS layers emit most hydrogen at
600\,\textcelsius\@.  After the hydrogen treatment, there is an additional
emission at 400\,\textcelsius\@.  Note that absolute peak heights bear a high
uncertainty in such measurements.  In particular, the fact that the original
peak at 600\,\textcelsius\ is still visible but slightly smaller after the
hydrogen treatment need not be a real effect.  However, the temperature axis is
accurate enough, and the peak separation large enough, to state that the
incorporated hydrogen is located and/or bonded differently from the hydrogen
that was already in the material.

IR absorption and effusion measurements give complementary information about
how the hydrogen is incorporated into the layer.  On the one hand, IR absorption
clearly shows that virtually the complete additional hydrogen is passivating
dangling bonds, which certainly is an encouraging result.  On the other hand,
effusion suggests that this hydrogen does not reach the innerst parts of the
layer.  Instead, it gets stuck close to the surface.  Note that ``surface''
includes here also inner surfaces of cracks and tunnels and structures of
enhanced hydrogen diffusion.

For determining the electrical conductivity, we evaporated two silver pad
contacts at a distance of 0.5~mm on the sample.  Then, we heat it in vacuum at
420~K for 30 minutes in order to evaporate most of the surface water film.  The
measurement itself is performed at room temperature using a Keithley~617
electrometer.  We measure the electrical current at voltages between $-100$ and
$+100$~V in order to detect non-ohmic behavior.  For photoconductivity
measurements, a xenon halogen lamp in conjunction with an infrared filter
(Schott KG\,7) provides the illumination.

Tab.~\ref{tab:layer-properties} includes the results of photo and dark
conductivity measurements.  They exhibit the most drastic change by hydrogen
treatment.  The dark conductivity decreases, which is advantageous for the use
in some types of devices.  But much more importantly, the photo conductivity
and the photo/dark ratio increase, the latter by a factor of almost~100.  All
of these changes make the material more suitable as an absorber material in
solar cells.

The defects and band tail states as discussed in the preceding paragraphs are
detrimental for electronic transport under illumination.  Defects are
recombination centres for charge carriers,\cite{abeles1982hydrogenated}
and band tail states are traps for them.\cite{tiedje1982band} This
explains the greatly increased photoconductivity after the hydrogen treatment.
For dark conductivity, however, the Fermi level moves away from the mobility
edge if less deep defects are present, so that the charge carrier balance is
fulfilled.  This causes a lower carrier concentration in the extended states
above the mobility edge, and therefore, decreases dark conductivity.

\begin{figure}
  \includegraphics[width=3in]{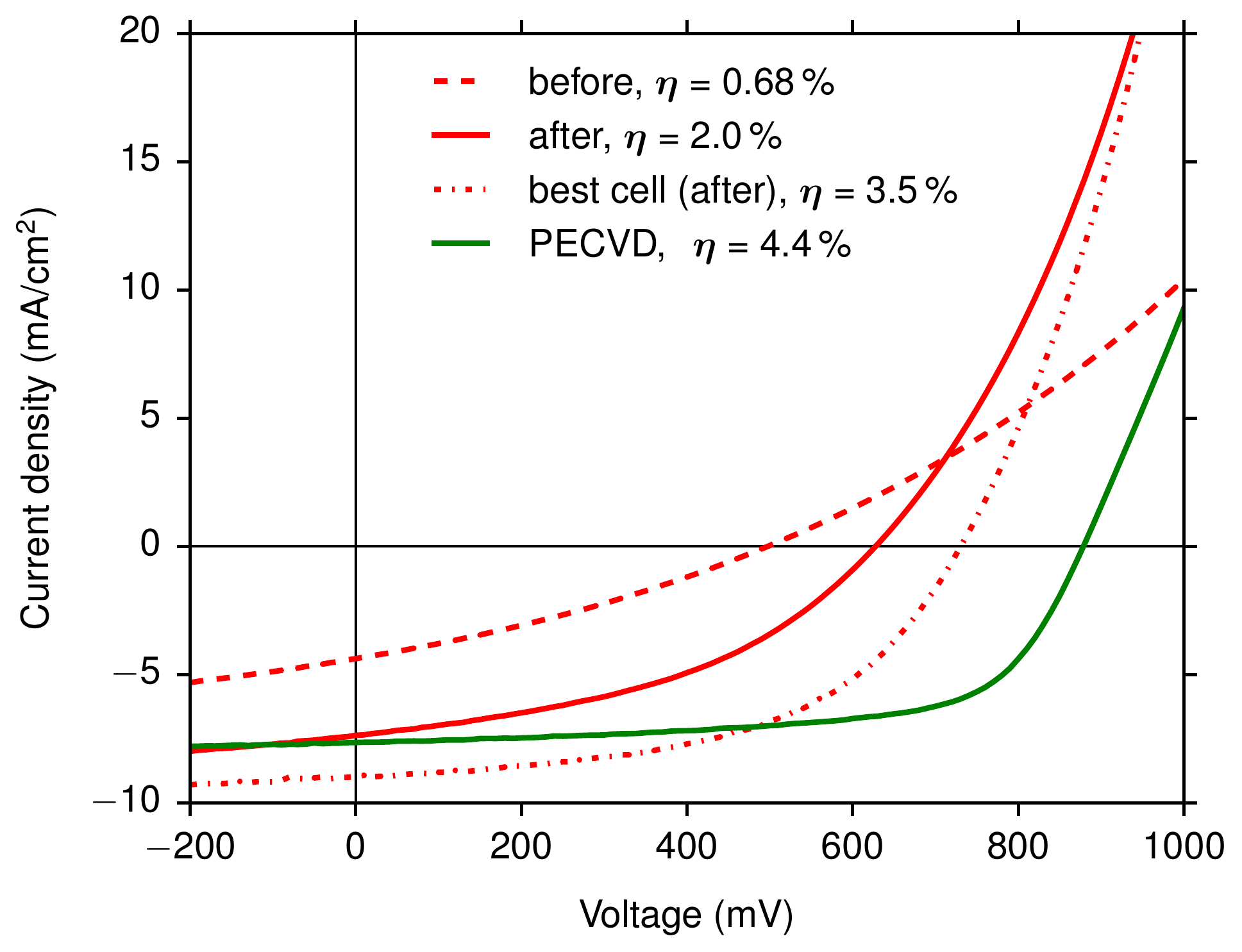}
  \caption{Characteristic IV curve before and after the hydrogen treatment.
    For comparison, we included the curve of our best cell device as published
    in Ref.~\onlinecite{bronger2014solution}, and a PECVD reference.}
  \label{fig:iv}
\end{figure}

We measure characteristic IV curves using a Keithley~238 source measure unit in
a sun simulator with AM\,1.5 spectrum (class~A), with a data point spacing of
10~mV\@.  During the measurement, we keep the sample at room temperature
$\pm1$\,\textcelsius\@.

Tab.~\ref{tab:cell-properties} and Fig.~\ref{fig:iv} show the influence of
the hydrogen treatment on solar cell devices.  Because the best cell published
in Ref.~\onlinecite{bronger2014solution} was not measured before passivation,
we present a comparison of a less efficient production run.  Nevertheless, the
best cell results are included for comparison, as are PECVD reference cell
results.  Note that the latter are adapted to the smaller thickness of the NPS
layers.  The details of this adaption can be found in
Ref.~\onlinecite{bronger2014solution}.

The hydrogen treatment significantly improves the cell characteristics in any
respect.  Most prominently, the series resistance (measured as the inverse
slope at the $V_{oc}$ point) drops by a factor of~2.6.  $I_{sc}$, $V_{oc}$, and
fill factor are increased by 68\,\%, 26\,\%, and 35\,\% respectively, and the
efficiency, which is a combination of these quantities, is tripled.

Note that the massive improvement in cell performance is attributed to the
hydrogen treatment alone, and thus is orthogonal to all other conceivable
optimization techniques (e.\,g.\ light trapping, thickness optimization,
conversion optimization, interface improvements).



A lack of electronic and optical quality in comparison to PECVD material
remains.  Both the density of defects and the widths of the band tails need to
be reduced to allow higher fill factor and open-circuit voltage.  Moreover,
only surface-near regions are passivated so far, which needs to be extended
into the bulk.

\bibliography{passivation}

\end{document}